# Observation of a Chiral Wave Function in Twofold Degenerate Quadruple Weyl System BaPtGe

Haoxiang Li[1,*], Tiantian Zhang[2,3,*,†], A. Said[4,*], Y. Fu[5], G. Fabbris[4], D. G. Mazzone[6], J. Zhang[1], J. Lapano[1], H. N. Lee[1], H. C. Lei[5,†], M. P. M. Dean[7], S. Murakami[2,3] and H. Miao[1,†]

[1]Material Science and Technology Division, Oak Ridge National Laboratory, Oak Ridge, Tennessee 37831, USA

[2]Department of Physics, Tokyo Institute of Technology, Okayama, Meguro-ku, Tokyo 152-8551, Japan

[3]Tokodai Institute for Element Strategy, Tokyo Institute of Technology, Nagatsuta, Midori-ku, Yokohama, Kanagawa 226-8503, Japan

[4]Advanced Photon Source, Argonne National Laboratory, Argonne, Illinois 60439, USA

[5]Department of Physics and Beijing Key Laboratory of Opto-Electronic Functional Materials and Micro-devices, Renmin University of China, Beijing, China

[6]Laboratory for Neutron Scattering and Imaging, Paul Scherrer Institut, CH-5232 Villigen, Switzerland

[7]Condensed Matter Physics and Materials Science Department, Brookhaven National Laboratory, Upton, New York 11973, USA

*These authors are contributed equally to this work.

†Correspondence should be addressed to ttzhang@stat.phys.titech.ac.jp, hlei@ruc.edu.cn and miaoh@ornl.gov.

**Topological states in quantum materials are defined by non-trivial topological invariants, such as the Chern number[1-5], which are properties of their bulk wave functions. A remarkable consequence of topological wave functions is the emergence of edge modes, a phenomenon known as bulk-edge correspondence, that gives rise to quantized or chiral physical properties[6-14]. While edge modes are widely presented as signatures of non-trivial topology, how bulk wave functions can manifest explicitly topological properties remains unresolved. Here, using high-resolution inelastic x-ray spectroscopy (IXS) combined with**

**first principles calculations, we report experimental signatures of chiral wave functions in the bulk phonon spectrum of BaPtGe, which we show to host a previously undiscovered twofold degenerate quadruple Weyl node. The chirality of the degenerate phononic wave function yields a non-trivial phonon dynamical structure factor, $S(Q,\omega)$, along high-symmetry directions, that is in excellent agreement with numerical and model calculations. Our results establish IXS as a powerful tool to uncover topological wave functions, providing a key missing ingredient in the study of topological quantum matter.**

In high-energy physics, chirality is an inherent quantum property of fundamental particles that originates from the Poincaré group. In condensed matter systems, however, chiral quasiparticles (QPs) can only emerge from low-energy effective models that must respect crystalline symmetries. Well-known examples are Weyl fermions in topological semimetals[7,9,12,13,15,16] and Weyl bosons in chiral lattices and optical and spin systems[14,17-24]. Although these chiral quasiparticles emerge from the material's bulk wave functions, existing experimental studies have focused either on the consequences of these quasiparticles, such as Fermi-arc surface states[25-34] and chiral anomalies[35-39], or argued based on a comparison of eigenvalues (band dispersions) between experiment and theory[21,22,40], leaving the bulk wave function unresolved. This is unfortunate as the information of topological invariants, such as the Chern Number, $C$, and the chirality of QPs, is carried by the bulk wave functions. Filling this gap is challenging as it requires a bulk sensitive spectroscopic probe that has high energy and momentum resolutions and a well-defined scattering cross-section that can be analytically or numerically calculated. In this letter, we demonstrate that high-resolution inelastic x-ray scattering (IXS) combined with first-principles calculations provides direct evidence of topological phononic wave functions in the chiral cubic crystal BaPtGe.

A fascinating property of BaPtGe is that it is predicted to host multiple chiral topological states, including a new and undiscovered twofold degenerate quadruple Weyl node (TQW)[41] that can only be realized in bosonic excitations, like phonons, or electronic structures without spin-orbit coupling. Past work has tended to assume that a Weyl node with large $C$ requires a large band degeneracy, $n$[16]. Indeed only $C \leq n$ has been experimentally observed to date. A remarkable feature of TQW is that a large Chern number, $C=4$, is realized in the smallest possible band degeneracy, $n=2$. This unusual topological structure is induced by the chiral, cubic and time-reversal symmetry of BaPtGe (space group P2$_1$3 #198) shown in Fig. 1a and d that enforce certain Weyl nodes at the time-reversal-invariant $\Gamma$ point. The low-energy effective Hamiltonian for the two modes that make up a TQW point is

$$H(\boldsymbol{q}) = \begin{pmatrix} Aq_xq_yq_z & B^*(q_x^2 + \omega^2 q_y^2 + \omega q_z^2) \\ B(q_x^2 + \omega q_y^2 + \omega^2 q_z^2) & -Aq_xq_yq_z \end{pmatrix} , \quad (1)$$

where $\omega = e^{2\pi i/3}$, $A$ is a real constant, and $B$ is a complex constant. $\boldsymbol{q}$ is the reduced momentum transfer in the first Brillouin zone. The leading $q^2$ term distinguishes TQW from all other Weyl nodes proposed in previous studies (see Fig. 1c)[16]. For instance, the two modes that arise from diagonalizing Eq. (1) have special pseudo-spin properties (see methods). Contours encircling the TQW point involve four windings of the pseudo-spin, reflecting the large $C=4$[41]. In Fig. 1d, we show pseudo-spin orientations at the eight R ($\pm 0.5, \pm 0.5, \pm 0.5$) points. At these high symmetry locations, pseudo-spin points along the $\pm z$-direction with its sign (marked as blue or yellow in Fig. 1d) determined by the sign of $q_xq_yq_z$. Consequently, the phonon wavefunctions of BaPtGe that make up the TQW point involve opposite out-of-phase chiral atomic motions around time-reversal related R points either side of the $\Gamma$-point. Beside the TQW, BaPtGe also features other topological structures, including the threefold double Weyl phonon with *n*=3 and *C*=2 at the Γ point and fourfold double Weyl phonon with *n*=4 and *C*=2 at the R point[14,21]. All of them have chiral bulk wave functions[14,21,24,41]. In Fig. 1e, we show density functional theory (DFT) calculation of the BaPtGe phonon dispersion along high symmetry directions. A low-energy TQW is identified near 6 meV at the Γ point (purple circle), which involves primarily atomic vibrations of the heaviest element Pt. The threefold and fourfold double Weyl nodes, that have been explored in both fermionic and bosonic systems[21,42], are marked by dashed squares at the Γ and R point, respectively.

We now illustrate how IXS probes the phononic wave functions. Under the harmonic approximation, the phonon excitations are described by

$$\sum_{d'} D_{dd'}(\mathbf{q}) e_d^j(\mathbf{q}) = \omega_{jq}^2 e_d^j(\mathbf{q}) \quad (2)$$

where $D(\mathbf{q})$ is the phonon dynamical matrix, and $\mathbf{q}$ is the reduced momentum transfer in the first Brillouin zone. $\omega_{jq}$ and $e_d^j(\mathbf{q})=[e_{dx}^j(\mathbf{q}), e_{dy}^j(\mathbf{q}), e_{dz}^j(\mathbf{q})]$ are the eigenvalue and wave function of phonon mode *j* at atom *d*, respectively. IXS measures the bulk phonon dynamical structure factor

$$S(\mathbf{Q} = \mathbf{q} + \mathbf{G}, \omega) = C_0 \sum_j \frac{1}{\omega_{qj}} |\sum_d \frac{f_d}{\sqrt{M_d}} \mathbf{Q} \cdot e_d^j(\mathbf{q}) e^{i\mathbf{Q}\cdot\mathbf{r}_d}|^2 \begin{Bmatrix} \langle n_{\omega_{qj}} + 1 \rangle \delta(\omega-\omega_{qj}) \\ \langle n_{\omega_{qj}} \rangle \delta(\omega+\omega_{qj}) \end{Bmatrix} \quad (3)$$

where **Q** and **G** are the total momentum transfer and the reciprocal lattice vector, respectively. $\langle n_{\omega_{qj}} \rangle$ is the Bose-Einstein distribution function. $f_d(\mathbf{Q})$ and $M_d$ are the atomic form factor and atomic mass of atom *d*, respectively. Wave function information enters $S(\mathbf{Q}, \omega)$ through the term $\mathbf{Q} \cdot e_d^j(\mathbf{q})$.

For a topological chiral wave function, the three components of $e_d^j(q)$ are complex numbers with different phase factors describing how atoms move out-of-phase with one another (see supplementary materials for movies of the chiral atomic motions). The chirality results in a non-trivial interference between the displacements of different atoms, which strongly modify $S(Q, \omega)$ and is encoded in large imaginary parts of $e_d^j(q)$. To quantitively show this effect, we simulate $S^{DFT}(Q, \omega)$, $R(Q, \omega)$ and $I(Q, \omega)$ along the high-symmetry direction $R_1$(-0.5, 2.5, 2.5)-$\Gamma$(0, 3, 3)-$R_2$(0.5, 3.5, 3.5) in Fig. 2a-2c, respectively. Here, $R(Q, \omega)$ and $I(Q, \omega)$ are calculations of the dynamical structure factor shown in Eq. (3), but with $e_d^j(q)$ replaced with its real part, $Re[e_d^j(q)]$, and imaginary part, $Im[e_d^j(q)]$, respectively. The chirality of the TQW near 6 meV is evidenced by the large intensity contribution from the imaginary component of the wave function along the $R_1$-$\Gamma$-$R_2$ direction. As we will discuss in detail later in the text, this observation is a direct consequence of the TQW Hamiltonian (Eq. 1).

To prove the existence of the topological chiral wave function in BaPtGe, we present experimental data of $S^{exp}(Q, \omega)$ that has been measured by IXS along the three high-symmetry directions, $X_1$(0, 3, 2.5)-$\Gamma$-$X_2$(0, 3, 3.5), $M_1$(0, 2.5, 2.5)-$\Gamma$-$M_2$(0, 3.5, 3.5) and $R_1$(-0.5, 2.5, 2.5)-$\Gamma$-$R_2$(0.5, 3.5, 3.5), in Fig. 3a-3c, respectively. In order to make a direct comparison with IXS, we show the experimental resolution-convoluted $S_r^{DFT}(Q, \omega)$ and $R_r(Q, \omega)$ along the same momentum directions in Fig. 3d-3f and Fig. 3g-3i. Here the subscript, $r$, stands for resolution to distinguish from $S^{DFT}(Q, \omega)$ and $R(Q, \omega)$. $S^{exp}(Q, \omega)$ and $S_r^{DFT}(Q, \omega)$ show excellent agreement, including the flatband near the TQW and the strong asymmetry of the threefold double Weyl optical and acoustic modes. This consistency is further confirmed by quantitative comparisons between experiment and theory shown in Fig. 4, where Fig. 4a-4c displays the extracted $Q$-dependent peak positions from $S^{exp}(\mathbf{Q},\omega)$ and $S_r^{DFT}(\mathbf{Q},\omega)$ and Fig. 4d-4f shows the extracted peak intensities near the TQW from $S^{exp}(\mathbf{Q},\omega)$, $S_r^{DFT}(\mathbf{Q},\omega)$ and $R_r(\mathbf{Q},\omega)$. Most interestingly, we find that wave functions near the TQW are primarily real along the $X_1$-$\Gamma$-$X_2$ and $M_1$-$\Gamma$-$M_2$ directions (Fig. 4d and 4e), in strong contrast to the $R_1$-$\Gamma$-$R_2$ direction (Fig. 4f), where the real-part of the wave function yields weak intensity near the TQW.

To understand the essential properties of the wave functions in different directions, we developed a simplified model that considers only Pt atoms in the unit cell, occupying the Wyckoff position $4a^{43}$ with locations $r_{Pt1}$=($c, c, c$), $r_{Pt2}$=($-c+\frac{1}{2}, -c, c+\frac{1}{2}$), $r_{Pt3}$=($-c, c+\frac{1}{2}, -c+\frac{1}{2}$), $r_{Pt4}$=($c+\frac{1}{2}, -c+\frac{1}{2}, -c$). At the $\Gamma$ point, group theory dictates that we can describe the atomic motions in terms of three non-degenerate basis states with irreducible representations of $\Gamma_1^{(1)}$, $\Gamma_2^{(1)}$, $\Gamma_3^{(1)}$, and three threefold degenerate basis states with irreducible representations of $\Gamma_4^{(3)43}$. The basis states of $\Gamma_2^{(1)}$ and $\Gamma_3^{(1)}$ that forms the TQW can be derived by imposing chiral cubic crystal symmetry and time-reversal symmetry:

$$\phi_{\Gamma_2} = \left(e_{Pt1}^{\Gamma_2}, e_{Pt2}^{\Gamma_2}, e_{Pt3}^{\Gamma_2}, e_{Pt4}^{\Gamma_2}\right) = (1, \omega, \omega^2, -1, -\omega, \omega^2, -1, \omega, -\omega^2, 1, -\omega, -\omega^2)/\sqrt{12} \quad (4)$$

$$\phi_{\Gamma_3} = \left(e_{Pt1}^{\Gamma_3}, e_{Pt2}^{\Gamma_3}, e_{Pt3}^{\Gamma_3}, e_{Pt4}^{\Gamma_3}\right) = (1, \omega^2, \omega, -1, -\omega^2, \omega, -1, \omega^2, -\omega, 1, -\omega^2, -\omega)/\sqrt{12} \quad (5)$$

where $\omega = e^{2\pi i/3}$. $\phi_{\Gamma_2}$ and $\phi_{\Gamma_3}$ have opposite chirality following $\phi_{\Gamma_2} = \phi_{\Gamma_3}^*$. In the vicinity of the $\Gamma$ point, the $q$-dependent wave functions can be obtained by diagonalizing Eq. (1) in terms of our $\phi_{\Gamma_2}$ and $\phi_{\Gamma_3}$ basis states. Along the [100] and [011] directions, the diagonal components of Eq. (1) are zero, imposing purely real wave functions, $\psi_1 = (e^{i\theta/2}\phi_{\Gamma_2} + e^{-i\theta/2}\phi_{\Gamma_3})/\sqrt{2}$ and $\psi_2 = -i(e^{i\theta/2}\phi_{\Gamma_2} - e^{-i\theta/2}\phi_{\Gamma_3})/\sqrt{2}$, where $\theta$=arg($B$). In contrast, along the [111] direction, the off-diagonal components are zero yielding $\psi_1 = \phi_{\Gamma_2}$ and $\psi_2 = \phi_{\Gamma_3}$, which have large imaginary components. This model analysis thus demonstrates that the directional wave function observed in Fig. 3 and 4, is a direct consequence of the TQW Hamiltonian. This feature is unique among all experimentally investigated topological structures (see supplementary materials). Figure 4g and 4h depict the chiral atomic motion of the TQW along the [111] direction. The results are calculated from the simplified model with the colored arrows representing the directions of Pt motions at a given time. The opposite chirality yields opposite Pt-motions, in agreement with our numerical calculation of BaPtGe (see supplementary materials for movies of the chiral atomic motions). Finally, we note that the intensity asymmetry of the threefold double Weyl modes is also due to chiral wave functions as directly evidenced from our numerical calculations (Fig. 2). However, this finite $q$ effect is beyond our simplified model where the effective Hamiltonian of threefold double Weyl contains only linear terms (see supplementary materials). Slightly away from the Γ point, higher order terms become important and the linear term approximation fails to describe the phonon excitations. For the TQW, the linear term is absent, and Eq. (1) with its square and cubic terms remains valid for an extend momentum space.

Our experimental discovery of the TQW in BaPtGe paves a new route to uncover topological wave functions. Similar attempts have been made in the study of topological electronic structures via angle-resolved photoemission spectroscopy (ARPES)[44-47]. While the momentum dependent ARPES intensity provides important clue on the symmetry of the surface wave function, the presence of surface Coulomb potentials makes it difficult to calculate the ARPES intensity from first principles[44]. Our approach, although based on IXS, can also be extended to inelastic neutron scattering to explore the bulk wave functions of topological phonons and magnons, which may play a crucial role in the giant thermal Hall effect observed in several correlated quantum materials[48,49].

**Methods:**

**Sample preparation and characterizations:** To grow BaPtGe single crystals, all the elements were stored and manipulated in an argon-filled glovebox with moisture and oxygen levels less than 0.1 ppm. Samples were prepared by the In-flux method. The high-purity Ba (lumps), Pt (powder), Ge (lumps), and In (grains) were put into a corundum crucible. The charged crucible and another catching crucible were sealed into a Ta tube with partially filled Ar gas. The Ta tube was sealed into quartz tube in order to avoid oxidation at high temperature. The tube was heated to 1180 °C in 12 h and held there for another 12 h. Then the ampoule was cooled down to 900 °C at a rate of ~ 2.5 °C/h. At this temperature, the flux was removed by centrifugation to obtain the shiny cubic-like crystals. The X-ray diffraction pattern of a BaPtGe single crystal was measured using a Bruker D8 X-ray diffractometer with Cu Kα radiation ($\lambda$ = 0.15418 nm) at room temperature. The elemental analysis was performed using energy-dispersive X-ray spectroscopy analysis in a FEI Nano 450 scanning electron microscope. No significant deviations from the expected structure and stoichiometry were observed in either measurement.

**Inelastic x-ray scattering:** The experiments were conducted at beam line 30-ID-C (HERIX) at the Advanced Photon Source (APS)[50]. The highly monochromatic x-ray beam of incident energy $E_i$ = 23.7 keV ($\lambda$ = 0.5226 Å) was focused on the sample with a beam cross section of ~35 × 15 μm² (horizontal × vertical). The overall energy resolution of the HERIX spectrometer was $\Delta E \sim$ 1.4 meV (full width at half maximum). The measurements were performed in transmission geometry. Typical counting times were in the range of 120 to 240 seconds per point in the energy scans at constant momentum transfer $Q$. H, K, L are defined in the cubic structure (space group #198 P2$_1$3) with a=b=c=6.747 Å at room temperature.

**Density functional theory calculation of phonon spectrum:** The phonon dispersions of BaPtGe were calculated using the density functional perturbation theory (DFPT) method and the Vienna Ab initio Simulation Package (VASP). The exchange-correlation potential was treated within the generalized gradient approximation (GGA) of the Perdew-Burke-Ernzerhof variety, where the kinetic energy cutoff was set to 400 eV. Integration for the Brillouin zone were done by using Monkhorst-Pack $k$-point grids which is equivalent to 8×8×9.

**Definition of pseudospin:** The Hamiltonian of the TQW (Eq. 1 in main text) can be written as $H_{2\times 2}(\boldsymbol{q}) = f_{i=x,y,z}(\boldsymbol{q}) \cdot \sigma_{i=x,y,z}$, where $\sigma_{x,y,z}$ are the Pauli matrices. This means that the wavefunctions can be characterized by their pseudospin, which is defined as $\hat{S}_i \equiv \frac{f_i(\boldsymbol{q})}{|f_i(\boldsymbol{q})|}$, where $i$ is a Cartesian index. This gives:

$$\hat{S}_x = \frac{f_x}{|f_x^2 + f_y^2 + f_z^2|} = \frac{q_z^2 - \frac{1}{2}(q_x^2 + q_y^2)}{|f_x^2 + f_y^2 + f_z^2|} \quad (7)$$

$$\hat{S}_y = \frac{f_y}{|f_x^2 + f_y^2 + f_z^2|} = \frac{\frac{\sqrt{3}}{2}(q_x^2 - q_y^2)}{|f_x^2 + f_y^2 + f_z^2|} \quad (8)$$

$$\hat{S}_z = \frac{f_z}{|f_x^2 + f_y^2 + f_z^2|} = \frac{q_x q_y q_z}{|f_x^2 + f_y^2 + f_z^2|} \quad . \quad (9)$$

Along each diagonal direction, such as the eight R-points, $\hat{S}_x = \hat{S}_y \equiv 0$, leaving only a single component of $\hat{S}_z = \pm 1$. Thus, pseudospins will point to $\pm q_z$ at eight R points, and the sign of $\hat{S}_z$ is only determined by the coordinate of the R point, i.e., the sign of $q_x q_y q_z$.

**Data availability:** The data that support the findings of this study are available from the corresponding author on reasonable request.

**Acknowledgements:** This research at Oak Ridge National Laboratory (ORNL) was sponsored by the Laboratory Directed Research and Development Program of ORNL, managed by UT-Battelle, LLC, for the U. S. Department of Energy (IXS experiment) and by the U.S. Department of Energy, Office of Science, Basic Energy Sciences, Materials Sciences and Engineering Division (IXS data analysis). Part of IXS data interpretation work at Brookhaven National Laboratory was supported by the U.S. DOE, Office of Science, Office of Basic Energy Sciences, Materials Sciences and Engineering Division under Contract No. DE- SC0012704 and by the U.S. DOE, Office of Basic Energy Sciences, Early Career Award Program under Award No. 1047478. This research used resources of the Advanced Photon Source, a U.S. Department of Energy (DOE) Office of Science User Facility, operated for the DOE Office of Science by Argonne National Laboratory under Contract No. DE-AC02-06CH11357. Extraordinary facility operations were supported in part by the DOE Office of Science through the National Virtual Biotechnology Laboratory, a consortium of DOE national laboratories focused on the response to COVID-19, with funding provided by the Coronavirus CARES Act. T.T. Z. and S. M. acknowledge the supports from Tokodai Institute for Element Strategy (TIES) funded by MEXT Elements Strategy Initiative to Form Core Research Center. S. M. also ac- knowledges support by JSPS KAKENHI Grant Number JP18H03678. H.C.L acknowledges the supports from the National Key R&D Program of China (Grant Nos. 2016YFA0300504, and 2018YFE0202600), the National Natural Science Foundation of China (Grant Nos. 11774423 and 11822412), the Fundamental Research Funds for the Central Universities, and the Research Funds of Renmin University of China (Grant Nos. 18XNLG14 and 19XNLG17).

**Figures:**

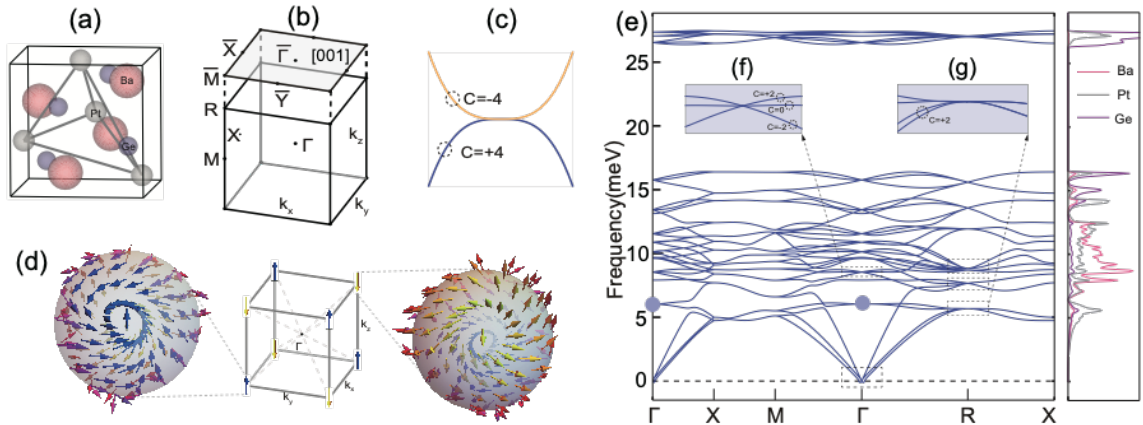

**Figure 1: Twofold degenerate quadruple Weyl node and chiral wave functions.** (a), (b) Crystal structure and Brillouin zone of BaPtGe, respectively. The high-symmetry points in the three-

dimensional BZ and the projected two-dimensional BZ are shown in black and light grey, respectively. (c) TQW node with $C = +4, -4$ for each band. (d) Pseudo-spin at the eight symmetry related R points. Colored arrows represent pseudo-spin directions. Zoomed pseudo-spin textures near time-reveal-related momenta (0.5,-0.5,-0.5) and (-0.5,0.5,0.5) in reciprocal lattice units (r.l.u.) are showing opposite chirality. (e) DFT calculated phonon dispersion of BaPtGe. Solid circles at the Γ points correspond to the theoretically predicted TQW. Dashed rectangles at the Γ and R points are corresponding to threefold and fourfold double Weyl nodes, which are zoomed in (f-g).

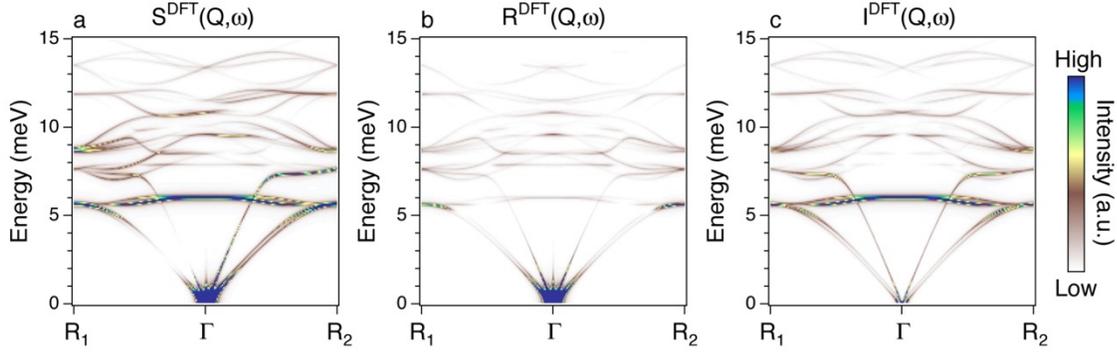

**Figure 2: DFT calculated dynamical structural factor.** **(a-c)** DFT calculated $S^{DFT}(\mathbf{Q},\omega), R(\mathbf{Q},\omega)$ and $I(\mathbf{Q},\omega)$ along the high-symmetry $R_1$(-0.5, 2.5, 2.5)-Γ(0, 3, 3)-$R_2$(0.5, 3.5, 3.5) direction, respectively. $R(\mathbf{Q},\omega)$ and $I(\mathbf{Q},\omega)$ considers only the real and imaginary part of $e_d^j(\mathbf{q})$. (a-c) are shown in the same color scale.

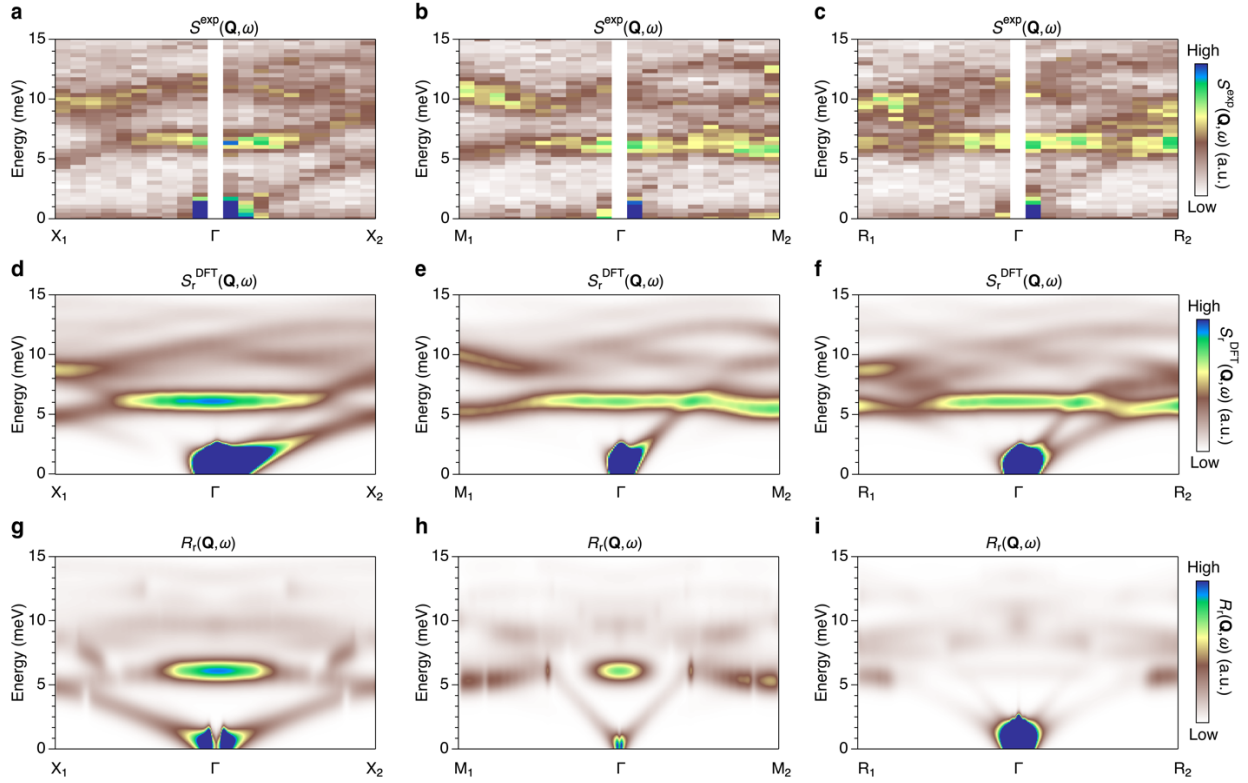

**Figure 3: Directional chiral wave function of the TQW. (a-c)** Experimental $S^{exp}(\mathbf{Q},\omega)$ results along the three high-symmetry directions [001], [011] and [111], where $X_1$= (0,3,2.5), $X_2$ =(0,3,3.5), $M_1$ =(0,2.5,2.5), $M_2$ =(0,3.5,3.5), $R_1$ =(-0.5,2.5,2.5), $R_2$ =(0.5,3.5,3.5). Experimental resolution convoluted $S_r^{DFT}(\mathbf{Q},\omega)$, $R_r(\mathbf{Q},\omega)$ are shown in **(d-f)** and **(g-i)**, respectively. The convolution function is determined by measuring the elastic scatting from plexiglass, which possesses a pseudo-voigt line shape with $\Delta E$~1.4 meV. (d-i) are shown in the same color scale.

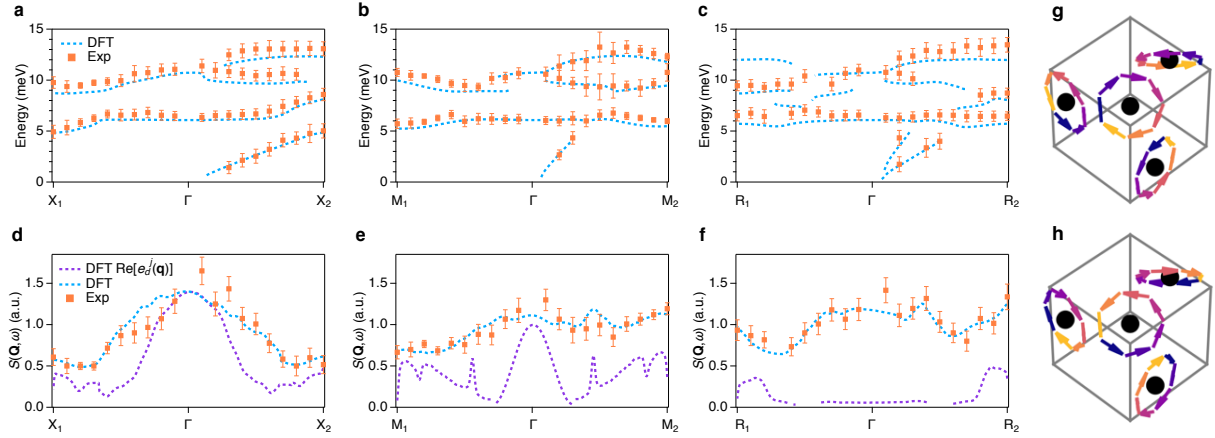

**Figure 4:** *Q*-dependent phonon intensity. (**a-c**) The extracted peak intensities from the experimentally determined $S^{exp}(\mathbf{Q},\omega)$ (orange dots) and calculated $S_r^{DFT}(\mathbf{Q},\omega)$ (blue dashed lines) along [001], [011] and [111] directions, respectively. The curve near 6 meV corresponds to the TQW. (**d-f**) *Q*-dependent peak intensities of the TQW that are extracted from $S^{exp}(\mathbf{Q},\omega)$, $S_r^{DFT}(\mathbf{Q},\omega)$, $R_r(\mathbf{Q},\omega)$, respectively. Vertical bars in panel (**a-c**) denote errors of the peak positions estimated based on the energy resolution and counting statistics. Error bars in panel (**d-f**) denote one standard deviation assuming Poissonian counting statistics in the measured $S(\mathbf{Q},\omega)$ intensity. (**g**) and (**h**) depict the chiral atomic motion of the TQW with *C*=+4 and -4, respectively. The results are calculated from the simplified model along the [111] direction. Colored arrows are pointing to the directions of Pt motions at a given time. Chirality is determined by the clockwise (**g**) and anti-clockwise (**h**) motions.